\documentclass[showpacs,twocolumn,aps,amsmath,floatfix]{revtex4}
\usepackage{epsf}
\begin{document}
\title{Finite Size Fluctuations in Interacting Particle Systems}
\author{E.~Ben-Naim}
\email{ebn@lanl.gov}
\affiliation{Theoretical
Division and Center for Nonlinear Studies, Los Alamos National
Laboratory, Los Alamos, New Mexico, 87545}
\author{P.~L.~Krapivsky}
\email{paulk@bu.edu} \affiliation{Center for Polymer Studies and
Department of Physics, Boston University, Boston, MA, 02215}

\begin{abstract}

  Fluctuations may govern the fate of an interacting particle system
  even on the mean-field level.  This is demonstrated via a three
  species cyclic trapping reaction with a large, yet finite number of
  particles, where the final number of particles $N_f$ scales
  logarithmically with the system size $N$, $N_f\sim\ln N$.
  Statistical fluctuations, that become significant as the number of
  particles diminishes, are responsible for this behavior. This
  phenomenon underlies a broad range of interacting particle systems
  including in particular multispecies annihilation processes.

\end{abstract}

 \maketitle

\section{Introduction}

Balls-in-boxes (urn) models provide a handy laboratory for studying
conceptual issues. The celebrated Erenfest model \cite{ee,ks,mk}, for
instance, has led to the reconciliation of the reversibility and the
recurrence paradoxes with Boltzmann's $H$ theorem.  Recent examples
include urn models of aging \cite{rf,gbm,gl} and of discretized
quantum gravity \cite{bbd}. Urn models have been studied extensively
in probability theory \cite{jk,kmr,fcp} and have found applications
ranging from biology \cite{ksd} to computer science \cite{bp,afp}.

Perhaps the most well-known manifestation of the role of fluctuations
in stochastic processes is the Eggenberger-P\'olya urn model
\cite{ep}. One starts with two marbles of different colors, draws a
marble randomly and puts it back together with another marble of the
same color. When the process is repeated indefinitely, the fraction of
marbles of a given color saturates at some limiting value. The
corresponding limit is a random variable that is uniformly distributed
between zero and one. Thus, initial fluctuations are locked in, a
striking example of the lack of self-averaging.

Inspired by this example, our goal is to quantify finite size
fluctuations in interacting particle systems using urn
models. Finite-size corrections are important because they govern for
example how a system converges to the thermodynamic limit, yet they
remain largely unresolved even in elementary stochastic processes
\cite{aal,jls,ve,kr,kr1,bk}.

We study the role of fluctuations using a three color cyclic urn
model.  Initially, the urn contains three different types of
balls. Then, two balls are picked randomly. If they are dissimilar,
following a cyclic rule, one of the balls is returned to the urn and
the other is removed from the system.  This urn model is different
from the Eggenberger-P\'olya type models in two ways. First, the
number of balls is {\it decreasing} rather than increasing. Second,
our model is {\it nonlinear} because picking two balls rather than one
implies that different type balls interact with each other.

We define the state of the system when one of the species is
depleted to be the final state. Starting from the natural initial
conditions where there are $N$ balls of each type, we ask: ``How many
balls are there in the final state?''. Our main result is that the
average number of balls in the final state $N_f$ scales
logarithmically with the system size $N$.

Statistical fluctuations are ultimately responsible for this
behavior. As long as the system contains enough particles (balls),
the average number of particles faithfully characterizes the state
of system. However, as particles deplete, the uncertainty with
respect to how many particles remain grows and moreover, it
governs how many particles are finally left.

This phenomenon and the mechanism underlying it are generic to
interacting particle systems with a decreasing number of
particles. We demonstrate this by considering multi-species
annihilation processes with $q$ species. In the three species model,
there is again a logarithmic enhancement of the variance in the number
of particles over the average number of particles, thereby leading to
a logarithmic growth law. In general, there is an algebraic
enhancement as long as the number of species is small enough, and
consequently, an algebraic growth law. Otherwise, when the number of
different species is large enough, statistical fluctuations are
negligible and the number of remaining particles is of order one. In
the most general case when $p$ balls are drawn from the urn, the
critical number of species is $q_c=2p-1$.

The rest of the paper is organized as follows. In Sec.~II, we give a
nontechnical presentation of the cyclic trapping model and highlight
the basic result. We then analyze the model in detail and obtain the
number density fluctuations by employing the $1/N$ expansion
(Sec.~III). The $q$-species annihilation model is treated in
Sec.~IV. We conclude with a few open questions (Sec. V).

\section{Cyclic Trapping Reactions}

Let us first define the model.  Initially, the urn contains balls of
three different types, denoted by A, B, and C. The balls interact via
a cyclic trapping reaction. Two balls are taken randomly out of the
box. If they are different, then one of the balls is returned to the
urn according to the cyclic rule
\begin{equation}
\label{rule-abc}
A+B\to B,\qquad B+C\to C,\qquad C+A\to A,
\end{equation}
and the second ball is discarded. This elemental step is repeated
until one of the species becomes extinct. This reaction scheme is
reminiscent of the Lotka-Volterra cyclic food chain (or 
rock-paper-scissors) model, widely used in ecology and game theory
\cite{ksd,jdm,rps,krfb,hs}.

The state of the system is fully specified by the number of particles
of each type in the urn: $n$, $m$, and $l$, corresponding to particles
of type $A$, $B$, and $C$. The dynamics is clearly mean-field: every
dissimilar pair of particles is equally likely to interact. Therefore,
the transition $(n,m,l)\to (n-1,m,l)$ occurs with probability
$nm/(nm+ml+ln)$ and similarly for the two other transitions.

We start with the initial condition $n=m=l=N$ and stop the process
when one species becomes extinct. Surprisingly, the number of balls in
the final state scales logarithmically with the system size:
\begin{equation}
\label{n-final}
N_f\sim\ln N.
\end{equation}
Numerical simulations are consistent with this behavior (Fig.~1).  We
verified numerically that the scale $\ln N$ fully characterizes the
final number of particles. The distribution of the final number of
particles approaches a (nontrivial) limiting distribution when the
final number of particles is normalized by $\ln N$. Thus, the final
number of particles in a non-self-averaging quantity.

The problem is essentially combinatorial, and in principle, it can be
addressed by weighing all possible histories with the appropriate
transition probabilities \cite{sr}. It proves fruitful, however, to
treat the process dynamically. Choosing the rate $nm/N$ for the
transition $(n,m,l)\to (n,m-1,l)$ is consistent with the above
microscopic rules. Moreover, it leads to $N$-independent dynamics in
the thermodynamic limit.

The number densities $a=\langle n\rangle/N$, $b=\langle m\rangle/M$,
and $c=\langle l\rangle/N$ evolve according to the rate equations
\begin{eqnarray}
\label{con-eq-abc} \frac{da}{dt}=-ab,\qquad
\frac{db}{dt}=-bc,\qquad \frac{dc}{dt}=-ca.
\end{eqnarray}
Since initially $a(0)=b(0)=c(0)=1$, the number densities remain
equal throughout the entire process $a(t)=b(t)=c(t)=\rho(t)$ with
\begin{eqnarray}
\label{con-sol-abc}
\rho(t)=(1+t)^{-1}.
\end{eqnarray}
Naively assuming that throughout the process, fluctuations in the
number density are much smaller than the mean leads to the conclusion
that the final number of particles is of the order unity, $N_f\sim
{\cal O}(1)$. The corresponding terminal time scales linearly with the
system size, $t_f\sim N$. Below, we show that this assumption does not
hold when the total number of particles becomes sufficiently small.

\begin{figure}[t]
\centerline{\epsfxsize=8cm\epsfbox{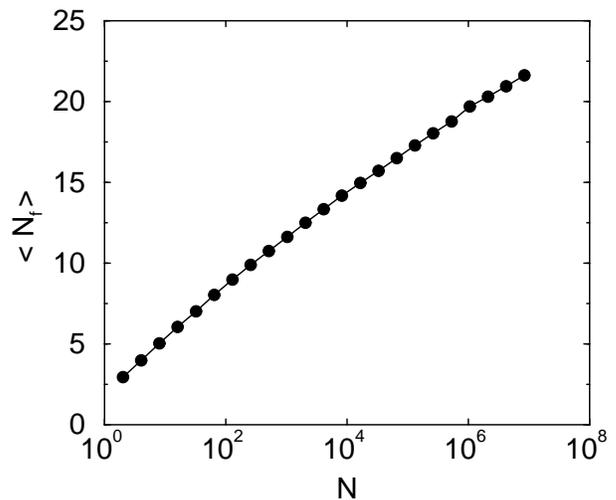}} \caption{The average
total number of particles in the final state as a function of the
system size. The data represents an average over $10^5$ realizations
of the cyclic trapping reaction process (\ref{rule-abc}).}
\end{figure}

The logarithmic growth in the number of particles can be deduced 
from the fluctuations in the number density. In the thermodynamic
limit, we expect that to leading order in $N$, both the total number
of particles and the variance in the number of particles are
proportional to the system size
\begin{eqnarray}
\label{sigma}
\langle n\rangle &\simeq &N\rho,\\
\langle n^2\rangle-\langle n\rangle^2&\simeq& N\sigma^2. \nonumber
\end{eqnarray}
We term $\sigma^2$ the intrinsic variance.  In Sec.~III, we shall 
utilize the van Kampen $1/N$ expansion \cite{ngv,br} to show that
asymptotically, the ratio between the intrinsic variance and the
density grows logarithmically with time
\begin{eqnarray}
\label{logt} \frac{\sigma^2}{\rho}\sim \ln t.
\end{eqnarray}
Thus, fluctuations eventually become larger then the density. Of
course, when they are comparable with the density, extinction is
possible. Hence, the number density (\ref{con-sol-abc}) characterizes
the particle number only up to a time scale $t_f$ obtained from the
validity criterion $N\rho(t_f)\sim \sqrt{N\sigma^2(t_f)}$. The terminal
time is therefore 
\begin{equation}
\label{t-final}
t_f\sim N(\ln N)^{-1}.
\end{equation}
Using $N_f\sim N\rho(t_f)$ we arrive at our main result
(\ref{n-final}).  Note that $\ln N$ is the leading contribution.  The
sub-leading contribution $\ln (\ln N)$ corresponding to nested
logarithms is tacitly ignored.

\section{Particle Number Fluctuations}

Fluctuations in the particle number are studied by expanding the
master equation in inverse powers of $N$ and keeping only the
leading order terms (large-$N$ expansion) \cite{ngv}.  The
probability $P(n,m,l;t)$ that the particle numbers are $n$, $m$,
and $l$ at time $t$ obeys the master equation
\begin{equation}
\label{distp-eq-abc}
\frac{d}{dt}P(n,m,l)=\left({\cal L}_{AB}+{\cal L}_{BC}+{\cal
L}_{CA}\right)P(n,m,l)
\end{equation}
with the initial condition
\hbox{$P_0(n,m,l)=\delta_{n,N}\delta_{m,N}\delta_{l,N}$}. The operator
${\cal L}_{AB}$ is
\begin{equation}
\label{lab-abc}
{\cal L}_{AB}P(n,m,l)=N^{-1}(\Delta_A-1)\left[nmP(n,m,l)\right];
\end{equation}
the operators ${\cal L}_{BC}$ and ${\cal L}_{CA}$ are defined via
similar formulas.  The difference operator $\Delta$ raises the
respective variable by one, e.g.
\begin{eqnarray}
\Delta_A\,f(n,m,l)=f(n+1,m,l).
\end{eqnarray}

Since in the thermodynamic limit, averages as well as variances grow
linearly with the system size as in Eq.~(\ref{sigma}), we introduce
the transformation $P(n,m,l)\to F(\alpha,\beta,\gamma)$ with
\begin{eqnarray*}
n=Na+N^{1/2}\alpha,\quad m=Nb+N^{1/2}\beta,\quad l=Nc+N^{1/2}\gamma.
\end{eqnarray*}
The intensive (random) variables $\alpha$, $\beta$, and $\gamma$ are
$N$-independent. These variables simply characterize fluctuations in
the respective particle numbers.

To find out how the distribution $F(\alpha,\beta,\gamma)$ evolves with
time, we write
\begin{equation}
\label{distf-eq-abc} F_t=({\cal M}_{AB}+{\cal M}_{BC}+{\cal
M}_{CA})F.
\end{equation}
It suffices to compute the evolution operator ${\cal M}_{AB}F$; the
two other operators are obtained by cyclic transposition. To obtain
the evolution operators, we replace the distribution $P$ by $F$ in
(\ref{distp-eq-abc}) and convert difference equations into
differential ones by expanding difference operators and keeping up to
second order terms, e.g., \hbox{$\Delta_A\to
1+\partial_n+\frac{1}{2}\partial_{nn}$}. Similarly, we replace
derivatives with respect to $n$ with derivatives with respect to
$\alpha$ using \hbox{$\partial_n=N^{-1/2}\partial_\alpha$}.  The time
derivative becomes \hbox{$\partial_t-N^{1/2}\dot
a\partial_\alpha-N^{1/2}\dot b\partial_\beta -N^{1/2}\dot
c\partial_\gamma$} where the overdot denotes differentiation with
respect to time. These transformations lead to 
\begin{widetext}
\begin{eqnarray}
\label{long}
F_t-N^{1/2}(\dot a F_\alpha+\dot b F_\beta+\dot c
F_\gamma)= (N^{-1}\partial_\alpha
+\frac{1}{2}N^{-1/2}\partial_{\alpha\alpha})
[(Na+N^{1/2}\alpha)(Nb+N^{1/2}\beta)F]+({\rm c.t.})
\end{eqnarray}
\end{widetext}
where (c.t.) denotes the two terms obtained by cyclic transposition of
the displayed term on the right-hand side. This master equation
contains terms of various orders in $N$. The order $N^{1/2}$ terms
vanish because the densities satisfy the rate equations
(\ref{con-eq-abc}). The next leading order term gives the evolution
operator
\begin{equation} \label{mab-abc} {\cal M}_{AB}F=
\rho\partial_\alpha[(\alpha+\beta)F]+\frac{1}{2}\rho^2 F_{\alpha\alpha}.
\end{equation}
Explicitly, the Fokker-Planck equation (\ref{distf-eq-abc}) is
\begin{eqnarray}
\label{master-abc}
F_t
&=&\rho[\partial_\alpha(\alpha+\beta)+\partial_\beta(\beta+\gamma)
+\partial_\gamma(\gamma+\alpha)]F\nonumber \\
&+&\frac{1}{2}\rho^2(F_{\alpha\alpha}+F_{\beta\beta}+F_{\gamma\gamma}).
\end{eqnarray}
This Fokker-Planck equation is subject to the initial condition
$F_0(\alpha,\beta,\gamma)=\delta(\alpha)\,\delta(\beta)\,\delta(\gamma)$.

Moments of the probability distribution $F(\alpha,\beta,\gamma)$
directly follow from (\ref{master-abc}). One simply multiplies
this Fokker-Planck equation by the desired powers of $\alpha$,
$\beta$, and $\gamma$, and integrates (by parts) with respect to
these three variables. Due to symmetry, there is essentially one
first moment: $\langle \alpha\rangle$; two second 
moments: $\langle \alpha^2\rangle, \langle\alpha\beta\rangle$;
three third moments: $\langle \alpha^3\rangle,
\langle\alpha^2\beta\rangle, \langle \alpha\beta\gamma\rangle$;
etc.  The first moment satisfies $\frac{d}{dt}\langle
\alpha\rangle =-2\rho\langle \alpha\rangle$ and since it vanishes
initially, $\langle \alpha\rangle=0$.  The two second moments 
are coupled
\begin{eqnarray}
\label{mom-abc} \frac{d\langle \alpha^2\rangle}{dt}
&=&-2\rho\langle\alpha^2\rangle-2\rho\langle\alpha\beta\rangle+\rho^2,\\
\frac{d\langle \alpha\beta\rangle}{dt}
&=&-\rho\langle\alpha^2\rangle-3\rho\langle\alpha\beta\rangle.\nonumber
\end{eqnarray}
These equations are inhomogeneous, so despite of the vanishing initial
conditions $\langle \alpha^2\rangle=\langle \alpha\beta\rangle=0$, the
solutions are non-trivial.

Writing $U=\langle \alpha^2\rangle + 2\langle \alpha\beta\rangle$ and
$V=\langle \alpha^2\rangle -\langle \alpha\beta\rangle$, we separate
the above equations
\begin{eqnarray}
\label{UV-eq-abc}
\frac{dU}{dt} &=&-4\rho U+\rho^2,\\
\frac{dV}{dt} &=&-\rho V+\rho^2.\nonumber
\end{eqnarray}
Using the number density (\ref{con-sol-abc}), we obtain the
explicit expressions
\begin{eqnarray}
\label{UV-sol-abc}
U&=&\frac{1}{3}\left[(1+t)^{-1}-(1+t)^{-4}\right]\,,\\
V&=&(1+t)^{-1}\ln(1+t).\nonumber
\end{eqnarray}
Physically, $U=\langle \alpha(\alpha+\beta+\gamma)\rangle$ quantifies
the correlation between the single particle number $n$ and the total
particle number $n+m+l$, while $V=\langle \alpha(\alpha-\beta)\rangle$
quantifies the correlation between the particle number $n$ and the
number difference $n-m$.  Intuitively, we expect that the quantity
$V$ is larger than $U$.  For a sufficiently large system, it may be
arbitrarily larger.

One of the two second moments is the intrinsic variance $\langle
\alpha^2\rangle\equiv \sigma^2$; explicitly,
\begin{eqnarray}
\label{var-sol-abc}
\frac{\sigma^2}{\rho}=\frac{2}{3}\left[\ln
(1+t)+\frac{1}{6}-\frac{1}{6}(1+t)^{-3}\right].
\end{eqnarray}
The other (normalized by the density) second moment quantifies
cross-correlations between different species numbers
\begin{eqnarray}
\label{cross-sol-abc}
\frac{\langle \alpha\beta\rangle}{\rho}
=-\frac{1}{3}\left[\ln
(1+t)-\frac{1}{3}+\frac{1}{3}(1+t)^{-3}\right].
\end{eqnarray}
The quantity $\langle \alpha\beta\rangle$ is always negative and
therefore, fluctuations between different particle numbers are
anti-correlated. Asymptotically, $\langle\alpha\beta\rangle\simeq
-\sigma^2/2$ with
\begin{eqnarray}
\label{var-lead-abc}
\sigma^2\simeq \frac{2}{3}\,t^{-1}\ln t.
\end{eqnarray}

Another important consequence of the structure of the Fokker-Planck
equation is that the multivariate distribution $P(n,m,l)$ is Gaussian
and fully characterized by the first and second order moments. This is
the case because the first order derivatives in (\ref{master-abc})
have linear coefficients \cite{ngv}. As a result, the individual
particle number distribution is also Gaussian
\begin{eqnarray}
\label{gaussian}
P(n,t)\simeq \frac{1}{\sqrt{2\pi N\sigma^2}}
\exp\left[-\frac{(n-N\rho)^2}{2N\sigma^2}\right].
\end{eqnarray}

\section{Multispecies Annihilation}

We have examined the question ``how many particles remain in the final
state?'' in a number of other interacting particle systems where
depletion or extinction occurs and found that generically, 
fluctuations play an important role. Using the same validity criterion,
and utilizing the van Kampen's $1/N$ expansion, one can determine the
final number of remaining particles.

We demonstrate this for multi-species annihilation processes.
Initially, the urn contains $q$ types of balls and the initial number
of each species is equal to $N$. For instance, when $q=3$,
\begin{equation}
\label{rule-q}
A+B\to 0,\qquad B+C\to 0,\qquad C+A\to 0.
\end{equation}
This process, introduced by ben-Avraham and Redner, was studied
primarily in low spatial dimensions via a number of numerical and
analytical techniques, yet it is still not fully understood
\cite{br,dht,zdb}.

The parameter $q$ is in principle integer. However, it is still
sensible to treat it as a continuous variable in the range
$2<q<\infty$. The $q$-species annihilation process can be reformulated
as a two-species annihilation model by combining $q-1$ of the species
into one group ($A$) and the remaining specie into a second group
($B$) \cite{bfk}. The reaction scheme becomes $A+A\to 0$ and $A+B\to
0$. The ratio between the reaction rates of the two channels,
$\frac{q-2}{q-1}$, is a continuous parameter that need not necessarily
correspond to an integer $q$.

The transition rates are as in the cyclic trapping reaction:
$(n,m,l,\ldots)\to (n-1,m-1,l,\ldots)$ occurs with rate $nm/N$.  For
symmetric initial conditions, the number density $\rho=a=b=c=\cdots$
satisfies
\begin{equation}
\label{con-eq-q} \frac{d\rho}{dt} =-(q-1)\rho^2,
\end{equation}
and the initial condition $\rho(0)=1$.  The concentration is therefore
\begin{equation}
\label{con-sol-q}
\rho=[1+(q-1)t]^{-1}.
\end{equation}

Fluctuations can be obtained following the same straightforward steps
the led to the evolution equations for the moments and we merely
highlight the derivation. For $q=3$, the probability distribution
$P(n,m,l)$ evolves according to (\ref{distp-eq-abc}) with the operator
${\cal L}_{AB}$ defined via
\begin{equation}
\label{def-lab} {\cal
L}_{AB}P=N^{-1}\left(\Delta_A\Delta_B-1\right)[nmP].
\end{equation}
The probability distribution $F(\alpha,\beta,\gamma)$ evolves
according to (\ref{distf-eq-abc}) with the evolution operator now
being
\begin{eqnarray}
\label{mab-q} {\cal
M}_{AB}F=\rho(\partial_\alpha+\partial_\beta)[(\alpha+\beta)F]+
\frac{\rho^2}{2}\left(\partial_{\alpha}
+\partial_{\beta}\right)^2F.\quad
\end{eqnarray}
For arbitrary $q$, there are $q(q-1)/2$ such operators. Again, the
first moment of $F$ vanish; the second moments are coupled as follows
\begin{eqnarray}
\label{abq} \frac{d\langle\alpha^2\rangle}{dt}
&=&-2(q-1)\rho\langle\alpha^2\rangle
-2(q-1)\rho\langle\alpha\beta\rangle+(q-1)\rho^2,\nonumber\\
\frac{d\langle\alpha\beta\rangle}{dt}&=&-2\rho\langle\alpha^2\rangle
-2(2q-3)\rho\langle\alpha\beta\rangle+\rho^2.
\end{eqnarray}
In contrast with the cyclic trapping reaction, the cross-correlation
initially grows, although asymptotically it is again negative.

Let $U=\langle\alpha^2\rangle+(q-1)\langle\alpha\beta\rangle$ and
$V=\langle\alpha^2\rangle-\langle\alpha\beta\rangle$; the former
quantity measures the correlation between $n$ and the total particle
number, the latter measures the correlation between $n$ and $n-m$. To
treat different values of $q$ on the same footing, we rescale the time
variable and introduce $\tau=(q-1)t$. The number density
(\ref{con-eq-q}) becomes $\rho=(1+\tau)^{-1}$ and the rate equations
for the quantities $U$ and $V$ are
\begin{eqnarray}
\label{UV-eqt-q}
\frac{dU}{d\tau}&=&-4\rho U+2\rho^2,\nonumber\\
\frac{dV}{d\tau}&=&-2\frac{q-2}{q-1}\rho V+\frac{q-2}{q-1}\rho^2.
\end{eqnarray}
The solution for the first quantity is therefore $q$-independent and
apart from the numerical prefactor, as in the cyclic trapping model,
\hbox{$U=\frac{2}{3}\left[(1+\tau)^{-1}-(1+\tau)^{-4}\right]$}.  Two
different behaviors are found for the second quantity:
\begin{eqnarray}
V=\begin{cases}
\frac{q-2}{q-3}\left[(1+\tau)^{-1}-(1+\tau)^{-2\frac{q-2}{q-1}}\right]
&q\neq 3\cr\frac{1}{2}(1+\tau)^{-1}\ln (1+\tau)&q=3\cr
\end{cases}
\end{eqnarray}

\begin{figure}[t]
\centerline{\epsfxsize=8cm\epsfbox{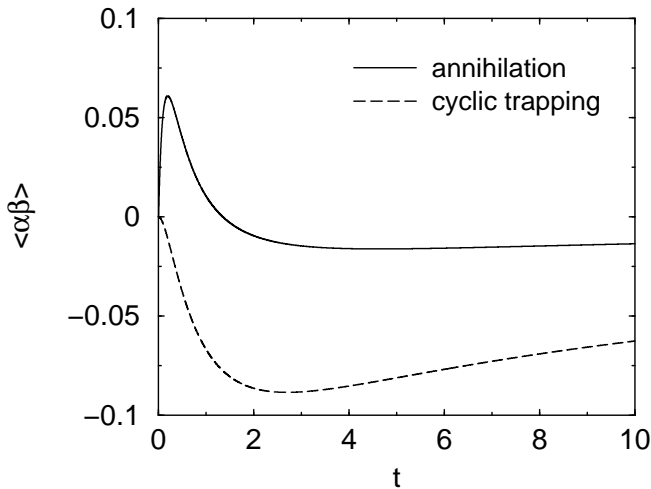}} \caption{The
cross-correlation versus time for the cyclic trapping model and the 
three-species annihilation.}
\end{figure}

Asymptotically, the cross-correlation is negative because
$\langle\alpha\beta\rangle\simeq -\frac{1}{q-2}V$ and so generically,
fluctuations between the numbers of different species are
anti-correlated. Early on, the cross-correlation increases,
but after a short transient it becomes negative (Fig.~2).

In the long time limit, $\sigma^2\simeq (1-q^{-1})V$:
\begin{eqnarray}
\label{sigma-q} \frac{\sigma^2}{\rho}\sim
\begin{cases}
t^{\frac{3-q}{q-1}}&q<3;\cr \ln t&q=3;\cr {\cal O}(1)&q>3.\cr
\end{cases}
\end{eqnarray}
Therefore, fluctuations are relevant asymptotically only when $q\leq
3$. Applying the criterion $\sqrt{N\sigma^2(t_f)}\sim N\rho(t_f)$
yields the final time $t_f(N)$ and consequently, the typical final
number of particles
\begin{eqnarray}
\label{nfinal-q}
N_f\sim
\begin{cases}
N^{(3-q)/2}&q<3;\cr
\ln N&q=3;\cr
{\cal O}(1)&q>3.\cr
\end{cases}
\end{eqnarray}
There is an algebraic growth in the fluctuations dominated regime
$q<3$. At the critical point $q_c=3$, logarithmic growth occurs.
Otherwise, the final number of particles saturates at a finite
value. Still, the final number diverges, $N_f\sim (q-3)^{-1}$, in the
vicinity of the critical point, $q\downarrow 3$. The saturation is
illustrated in Fig.~3 using numerical simulation data for the $q=4$
case.

\begin{figure}[t]
\centerline{\epsfxsize=8cm\epsfbox{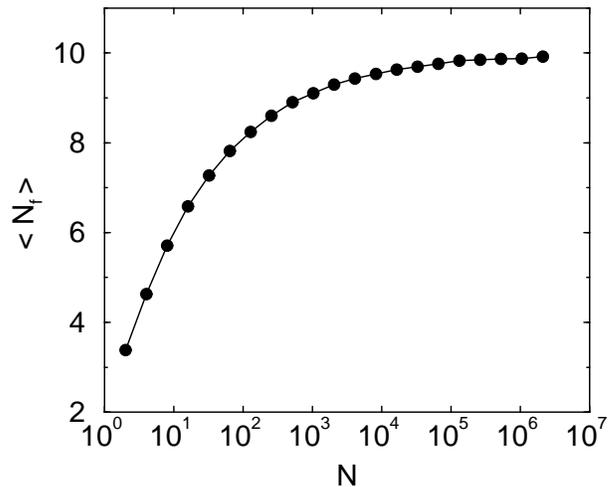}} \caption{The average
total number of particles in the final state as a function of the
system size for $q=4$. The data represents an average over $10^5$
realizations of the four-species annihilation process.}
\end{figure}

The case $q=2$ is special since there is a conservation law ($n-m$ is
conserved) and therefore, $V=0$. Consequently, \hbox{$\sigma^2\sim
\rho \sim t^{-1}$}.  If a $q$-species aggregation, rather than
annihilation process is considered, that is, when $A$ and $B$ interact
the outcome is either $A+B\to A$ or $A+B\to B$ (both taken with equal
probability), then this anomaly disappears \cite{br,dht} and
Eq.~(\ref{nfinal-q}) holds for $q=2$ as well.

One may ask ``why is the critical number of species equal to 3?''.
Mathematically, the answer is ultimately related to the smaller
eigenvalue of the $2\times 2$ matrix coupling the second moments. Yet
given the simplicity of the multispecies urn model, a heuristic and
more illuminating derivation may be possible after all. Finding such
an argument is an interesting challenge.

In this context, we mention a generalization of the urn model from
two-particle to the many-particle interactions. That is, instead of
picking $2$ particles, $p$ particles are picked and if they are all of
different species, all are removed from the system (this process is
well defined for $p\leq q$). The rate equations for the second moments
yield the critical number of species (see Appendix)
\begin{equation}
q_c(p)=2p-1.
\end{equation}
Thus, for ternary interactions $q_c=5$.  The structure of the phase
transition remains the same. There is an algebraic growth in the total
number of remaining particles as a function of the system size when
$q<q_c$, a logarithmic growth at the critical point $q=q_c$, and
saturation above the critical point $q>q_c$.  Last, we mention that a
similar phase transition underlies two-species reaction processes of
the type constructed from the $q$-species annihilation by separating 
species into two groups. In this case, although the transition depends
on the reaction rates rather than the number of species, its structure
remains the same.

\section{conclusion}

In summary, we considered interacting particle systems undergoing
depletion on the mean-field level. We showed that finite size
fluctuations display a rich behavior. The behavior is rather generic
and applies to a wide class of stochastic processes with a decreasing
number of particles. Typically, there is a phase transition as a
function of the number of species or the reaction rates. In one region
of parameter space, the final number of particles grows algebraically,
and in the other, it saturates at a finite value. The critical case is
marked by a logarithmic growth.  We conclude that the final number of
particles as a function of system size provides a practical probe of
statistical fluctuations.

The findings in the cyclic trapping model have a neat game theoretic
implication. In a rock-paper-scissor game involving fixed strategy
players and loser-is-out rules, the game ends when all remaining
players have the same strategy. If players pair randomly, then the
ultimate number of winners scales logarithmically with the total
number of players. Intuitively, we expect that a similar law emerges
for tournaments with simultaneous play.

Several questions arise naturally. Can one explain the critical number
of species using heuristic arguments? Statistical properties of the
final state and how the system approaches it are interesting as
well. For example, what is the number distribution of remaining
particles? How different are statistical properties of the system at
the very end of the process when only a single species remains?  We
observed that the convergence to the asymptotic behavior is much
faster when the first extinction occurs compared with the very end
state when only a single species remains.  Last, an interesting
question involves extremal characterization \cite{kr1,bk}: What is the
probability that one of the species is always the most or least
numerous?

We studied systems undergoing depletion. However, there are processes
in which depletion is possible but not certain, for example, infection
processes \cite{jdm}. It will be interesting to investigate finite
size fluctuations in this related class of interacting particle
systems.

\medskip

We are thankful to Matt Hasting, Sid Redner, and Zoltan Toroczkai for
useful discussions.  This research was supported by
DOE(W-7405-ENG-36).

\newpage
\appendix
\begin{widetext}
\section{$p$-particle annihilation}

For $p$-particle annihilation, the density evolves according to 
\begin{equation}
\frac{d}{dt}\rho=-{q-1\choose p-1}\rho^p.
\end{equation}
The evolution operators for the Fokker-Planck equation for $F$ are
straightforward generalizations of Eq.~(\ref{mab-q}). For example, for
the ternary ($p=3$) annihilation process $A+B+C\to 0$,
\begin{eqnarray}
{\cal M}_{ABC}F=\rho^{2}(\partial_\alpha+\partial_\beta+\partial_\gamma)[(\alpha+\beta+\gamma)F]+
\frac{\rho^3}{2}\left(\partial_{\alpha}+\partial_{\beta}+\partial_{\gamma}\right)^2F.\qquad
\end{eqnarray}
The second moments are coupled as follows 
\begin{eqnarray}
\frac{d}{dt}
\begin{pmatrix}\langle\alpha^2\rangle\cr \langle\alpha\beta\rangle\cr\end{pmatrix}
=-2\rho^{p-1}
\begin{pmatrix}
{q-1\choose p-1}& (p-1){q-1\choose p-1} \\
{q-2\choose p-2}& (p-1){q-2\choose p-2}+p{q-2\choose p-1}\\
\end{pmatrix}
\begin{pmatrix}\langle\alpha^2\rangle\cr \langle\alpha\beta\rangle\cr\end{pmatrix}
+\rho^p
\begin{pmatrix}
{q-1\choose p-1}\cr {q-2\choose p-2}
\end{pmatrix}.
\end{eqnarray}
Introducing the time variable $\tau=(p-1){q-1\choose p-1}t$, the
density is simply $\rho=(1+\tau)^{-1/(p-1)}$. The quantity
\hbox{$U=\langle\alpha^2\rangle+(q-1)\langle\alpha\beta\rangle$} satisfies
\hbox{$\frac{d}{d\tau}U+\frac{2p}{p-1}\rho^{p-1}U=\frac{p}{p-1}\rho^p$} and 
the solution is again $q$-independent 
\begin{equation}
U(\tau)=\frac{p}{2p-1}
\left[(1+\tau)^{-\frac{1}{p-1}}-(1+\tau)^{-\frac{2p}{p-1}}\right].
\end{equation}
The quantity \hbox{$V=\langle\alpha^2\rangle-\langle\alpha\beta\rangle$} satisfies 
\hbox{$\frac{d}{d\tau}V+\frac{2(q-p)}{(p-1)(q-1)}\rho^{p-1}V=\frac{q-p}{(p-1)(q-1)}\rho^p$}. The solution reads 
\begin{equation}
V(\tau)=
\begin{cases}
\frac{q-p}{q-(2p-1)}
\left[(1+\tau)^{-\frac{1}{p-1}}-(1+\tau)^{-\frac{2(q-p)}{(p-1)(q-1)}}\right]&q\neq 2p-1;\\
\frac{1}{2(p-1)}(1+\tau)^{-\frac{1}{p-1}}\ln (1+\tau)&q=2p-1.
\end{cases}
\end{equation}
Interestingly, in the fluctuation dominated regime, $q<2p-1$, the
exponent governing the terminal time is $p$-independent, $t_f\sim
N^{\frac{q-1}{2}}$.  The final number of particles is
\begin{eqnarray}
N_f\sim
\begin{cases}
N^{\frac{2p-1-q}{2(p-1)}}&q<2p-1;\\
\ln N&q=2p-1;\\
{\cal O}(1)&q>2p-1.
\end{cases}
\end{eqnarray}
\newpage
\end{widetext}


\begin{thebibliography}{99}

\bibitem{ee}
      P.~Erenfest and T.~Erenfest,
      Phys.\ Zeit. {\bf 8}, 311 (1907);
      P.~Erenfest and T.~Erenfest, Enzyklop\"adie der Mathematischen
      Wissenschaften, Vol.~IV, pt 32 (Berlin, 1911).

\bibitem{ks}
      F.~Kohlrausch and E.~Schr\"odinger,
      Phys.\ Zeit. {\bf 27}, 306 (1926).

\bibitem{mk}
      M.~Kac, {\em Probability and Related Topics in Physical Sciences}
      (Interscience Publishers, London, 1959).

\bibitem{rf}
      F.~Ritort, Phys.\ Rev.\ Lett. {\bf 75}, 1190 (1995);
      S.~Franz and  F.~Ritort, J. Phys. A {\bf 30}, L359 (1997).

\bibitem{gbm}
      C.~Godr\`eche, J.~P.~Bouchaud, and M.~Mezard,
      J. Phys. A {\bf 28}, L603 (1995).

\bibitem{gl}
      C.~Godr\`eche and J.~M.~Luck,
      J. Phys. A {\bf 29}, 1915 (1996); {\it ibid}
      {\bf 30}, 6245 (1997); {\it ibid} {\bf 32}, 6033 (1999).

\bibitem{bbd}
      P.~Bialas, Z.~Burda, and D.~Johnson, Nucl. Phys. B
      {\bf 493}, 505 (1997); {\it ibid} {\bf 542}, 413 (1999).

\bibitem{jk}
      N.~L.~Johnson and S.~Kotz, {\em Urn Models and their Applications}
      (Wiley, New York, 1977).

\bibitem{kmr} S.~Kotz, H.~Mahmoud, and P.~Robert, Stat.\ Probab.\
      Lett. {\bf 49}, 163 (2000).

\bibitem{fcp}
      P.~Flajolet, J.~Cabarro, and H.~Pekari, Ann.\ Prob. (2004).

\bibitem{ksd}
      K.~Sigmund, {\em Games of Life}
      (Oxford University Press, Oxford, 1993).

\bibitem{bp}
      A.~Bagchi and A.~K.~Pal, SIAM J. Alg.\ Disc.\ Meth. {\bf 6}, 394
      (1985).

\bibitem{afp}
      D.~Aldous, B.~Flannery, and J.~L.~Palacios,
      Prob.\ Eng.\ Infor.\ Sci. {\bf 2}, 293 (1988).

\bibitem{ep}
      F.~Eggenberger and G.~P\'olya, Zeit.\ Angew.\ Math.\ Mech. {\bf 3}, 279
      (1923).

\bibitem{aal}
      A.~A.~Lushnikov,
      J.\ Colloid.\ Inter.\ Sci.\ {\bf 65}, 276 (1977).

\bibitem{jls}
      J.~L.~Spouge,
      J.\ Colloid.\ Inter.\ Sci.\ {\bf 107}, 38 (1985).

\bibitem{ve}
      P.~G.~J. van Dongen and M.~H.~Ernst,
      J. Stat. Phys. {\bf 49}, 879 (1987).

\bibitem{kr}
      P.~L.~Krapivsky and S.~Redner,
      J.\ Phys.\ A {\bf 35}, 9517 (2002).

\bibitem{kr1}
      P.~L.~Krapivsky and S.~Redner,
      Phys. Rev. Lett. {\bf 89} 8703 (2002). 

\bibitem{bk}
      E.~Ben-Naim and P.~L.~Krapivsky,
      Europhys. Lett., in press (2003).

\bibitem{jdm}
     J.~D.~Murray, {\em Mathematical Biology}
     (Springer-Verlag, New York, 1989).

\bibitem{rps} The official rock-paper-scissors web page:
     http://www.worldrps.com/.

\bibitem{krfb}
      B.~Kerr, M.~A.~Riley, M.~W.~Feldman, and B.~J.~M.~Bhannan, 
      Nature {\bf 418}, 171 (2002).

\bibitem{hs}
      J.~Hofbauer and K.~Sigmund,
      Bull. Amer. Math. Soc. {\bf 40}, 479 (2003).

\bibitem{sr}
      S.~Redner,
     {\em A Guide to Fisrt Passage Processes}
     (Cambridge, New York, 2001).

\bibitem{ngv}
      N.~G.~Van Kampen, {\em Stochastic Processes in Physics and Chemistry}
      (North Holland, Amsterdam, 2003).

\bibitem{br}
      D.~ben-Avraham and S.~Redner, Phys.\ Rev.\ A {\bf 34}, 501 (1986).

\bibitem{dht}
      O.~Deloubriere, H.~J.~Hilhorst, and U.~C.~Tauber,
      Phys.\ Rev.\ Lett. {\bf 89}, 250601 (2002).

\bibitem{zdb}
      D.~X.~Zhong, R.~Dawkins, and D.~ben-Avraham,
      Phys.\ Rev.\ E {\bf 67}, 040101 (2003).

\bibitem{bfk}
      E.~Ben-Naim, L.~Frachebourg, and P.~L.~Krapivsky,
      Phys.\ Rev.\ E {\bf 53}, 3078 (1996).

\end{thebibliography}
\end{document}